\documentstyle[]{article}

\def\func#1{\mbox{\rm $#1$}}
\def\text#1{#1}
\font\Bbb=msbm10

\begin{document}

\title{QUANTUM EFFECTS FOR EXTRINSIC GEOMETRY OF STRINGS VIA THE GENERALIZED
WEIERSTRASS REPRESENTATION}
\author{ 
B. G. Konopelchenko and G. Landolfi \\
{\em
Dipartimento di Fisica, Universit\`a di Lecce, 73100, Lecce, Italy} \\
and \\
{\em
I.N.F.N., Sezione di Lecce, 73100, Lecce, Italy} 
}
\date{}
\maketitle{}

\begin{abstract}
The generalized Weierstrass representation for surfaces in $\Bbb{R}^{3}$ is
used to study quantum effects for strings governed by Polyakov-Nambu-Goto
action. Correlators of primary fields are calculated exactly in one-loop
approximation for the pure extrinsic Polyakov action. Geometrical meaning of
infrared singularity is discussed. The Nambu-Goto and spontaneous curvature
actions are treated perturbatively.
\end{abstract}

\section{Introduction}

Since the Polyakov's suggestion [1] to add new rigidity term (Polyakov
extrinsic action) to the old Nambu-Goto action, the string theory based on
such an extended action has been intensively studied (see \textit{e.g.}
[2]-[4]). Effects generated by an extrinsic geometry term have been analyzed
by several authors [5]-[13]. A canonical description of the strings
world-sheets by means of coordinates $\overrightarrow{X}$ of the target
space in which strings are assumed to evolve was not very successful. This
is mainly due to the presence of fourth order derivatives of $%
\overrightarrow{X}$ with respect to local coordinates on world-sheet. In
last years, the discretization of surfaces and associated matrix models have
been the most favorite tool to treat the problem (see \textit{e.g.} [14]). A
different approach has been developed in the papers [12]-[13]. It is based
on the description of the strings world-sheets via the Gauss map for
surfaces conformally immersed into the Euclidean spaces. Within this
approach both the Nambu-Goto and Polyakov actions can be written in terms of
constrained K\"{a}hler $\sigma $-model action. The use of the Gauss map
makes easier the calculations of quantum effects induced by extrinsic
geometry [12]-[13]. Unfortunately nonlinear constraints associated with the
Gauss map for nonminimal surfaces give rise to serious computational
difficulties.

In this paper we present an approach based on the generalized Weierstrass
representation introduced in [15]. Any surface in $\Bbb{R}^{3}$ can be
generated via this representation provided a system of two linear equations
is solved. Within the generalized Weierstrass representation the Nambu-Goto
action and, particularly, the Polyakov action have a very simple form.
This allows us to calculate the one loop correction to the background for
the full Polyakov action exactly. The calculations in the momentum space
occur to be convenient. The propagators of fields are found and their
infrared behavior is analyzed. Quantum correction to the classical
Nambu-Goto and spontaneous curvature actions are evaluated perturbatively.

Note that the generalized Weierstrass representation has been used recently
within a different technique [16]-[17] to evaluate quantum effects in the
string theory.

The paper is organized as follows. In section 2 we briefly summarize the
generalized Weierstrass representation for surfaces in $\Bbb{R}^{3}$. The
one-loop quantum effects of strings are studied in section 3. The Nambu-Goto
and spontaneous curvature actions are discussed in section 4. Section 5 is
devoted to discussion. 

\section{Generalized Weierstrass representation}

The generalized Weierstrass representation for a surface conformally
immersed in $\Bbb{R}^{3}$ ($\overrightarrow{X}(z,\overline{z}):\Bbb{C}%
\rightarrow \Bbb{R}^{3}$) is given by the formulae [15] 
\begin{equation}
\begin{array}{l}
X^{1}+iX^{2}=i\int_{\Gamma }\left( \overline{\psi }^{2}dz^{\prime }-%
\overline{\varphi }^{2}d\overline{z}^{\prime }\right) \quad , \\ 
X^{3}=-\int_{\Gamma }\left( \varphi \overline{\psi }dz^{\prime }+\psi 
\overline{\varphi }d\overline{z}^{\prime }\right) 
\end{array}
\label{2.1}
\end{equation}
where $z,\overline{z}$ $\in \Bbb{C}$, bar means complex conjugation, $\Gamma 
$ is a contour in $\Bbb{C}$ and complex-valued functions $\psi $, $\varphi $
obey the system of equations 
\begin{equation}
\partial \psi =p\varphi \qquad ,\qquad \overline{\partial }\varphi =-p\psi 
\label{2.2}
\end{equation}
where $p=p\left( z,\overline{z}\right) $ a real-valued function. The
formulae (\ref{2.1})-(\ref{2.2}) define a conformal immersion of a surface
into $\Bbb{R}^{3}$ with the induced metric 
\begin{equation}
ds^{2}=\left( \left| \psi \right| ^{2}+\left| \varphi \right| ^{2}\right)
^{2}dz\,d\overline{z}\qquad .  \label{2.3}
\end{equation}
The Gaussian and the mean curvature are 
\begin{eqnarray}
K &=&-\frac{4}{\left( \left| \psi \right| ^{2}+\left| \varphi \right|
^{2}\right) ^{2}}\left[ \log \left( \left| \psi \right| ^{2}+\left| \varphi
\right| ^{2}\right) \right] _{z\,\overline{z}}\quad ,  \nonumber \\
H &=&2\frac{p}{\left( \left| \psi \right| ^{2}+\left| \varphi \right|
^{2}\right) }\qquad .  \label{2.4}
\end{eqnarray}
Any surface in $\Bbb{R}^{3}$ can be represented in the form (\ref{2.1})-(\ref
{2.2}) [15], [18]. At $p=0$ one gets a minimal surface ($H=0$) and the
formulae (\ref{2.1})-(\ref{2.2}) are reduced to the classical Weierstrass
formulae for minimal surfaces.

The generalized Weierstrass representation (\ref{2.1})-(\ref{2.2}) is
equivalent to another Weierstrass type representation which was proposed in
[19] and has been used within the Gauss map approach in the papers
[12]-[13]. The equivalence is established by simple formulae [20] 
\begin{equation}
f=i\frac{\overline{\psi }}{\varphi }\qquad ,\qquad \eta =\frac{i}{2}\varphi
^{2}  \label{2.5}
\end{equation}
which however convert the linear system (\ref{2.2}) into a nonlinear one
which appeared in [19] and, consequently, in [12]-[13]. The fact that in the
generalized Weierstrass representation (\ref{2.1})-(\ref{2.2}) the functions 
$\psi $ and $\varphi $ are constrained by the linear equation (\ref{2.2})
(instead of a nonlinear one of [18]) is one of its advantages. One more
advantage is that the extrinsic Polyakov action $S_{P}=\int H^{2}\left[
dS\right] $ , where $[dS]$ is the area element, takes a very simple form
[20]. Indeed, the use of (\ref{2.4}) gives 
\begin{equation}
S_{P}=4\int p^{2}[d^{2}z]  \label{2.6}
\end{equation}
where $[d^{2}z]=\left( i/2\right) \,dzd\overline{z}$. 
The Nambu-Goto action $%
S_{NG}=\alpha _{0}\int \left[ dS\right] $ becomes 
\begin{equation}
S_{NG}=\alpha _{0}\int \left( \left| \psi \right| ^{2}+\left| \varphi
\right| ^{2}\right) ^{2}[d^{2}z]\qquad .  \label{2.7}
\end{equation}

The generalized Weierstrass representation (\ref{2.1})-(\ref{2.2}) has
allowed already to obtain several interesting results in differential
geometry of surfaces where the extrinsic Polyakov action is known for a long
time as the Willmore functional [21] (see [22]-[24]), in the theory of
liquid membranes [25] and in the string theory [26], [16]-[17]. The
representation (\ref{2.1})-(\ref{2.2}) gives also the possibility to define
an infinite class of integrable deformations of surfaces generated by the
modified Veselov-Novikov hierarchy [15]. A characteristic feature of these
deformations is that they preserve the extrinsic Polyakov action [20], [22].
This circumstance has been used in [16]-[17] to quantize the Willmore
surface (surfaces which provide extremum to the Willmore functional
(Polyakov action)).

The generalized Weierstrass representation (\ref{2.1})-(\ref{2.2}) can be
viewed as a parametrization of a surface in $\Bbb{R}^{3}$ in terms of $p$, 
$\psi $, $\varphi $. We will see that this parametrization is quite
convenient.

\section{ One loop effects}

One-loop corrections for the Polyakov action have been already studied in
[1], [12]-[13]. However nonlinear constraints associated with the Gauss
map did not allow to calculate one-loop corrections for the full Polyakov
action. In this section we will show that the generalized Weierstrass
representation enables to overcome this difficulty and provides us a deeper
geometrical understanding of results.

We will follow to the method of calculations proposed in [1] and then used
in [12]-[13]. So we start with the classical action 
\begin{equation}
S=\alpha _{0} \int \left( \left| \psi \right| ^{2}+\left| \varphi
\right| ^{2}\right) ^{2}\left[ d^{2}z\right] +\beta _{0}^{\prime }\int
\,p^{2}\left[ d^{2}z\right]  \label{3.1}
\end{equation}
where $\alpha _{0}$ and $\beta _{0}^{\prime }$ are the tension and extrinsic
coupling constant respectively.

The first step is to take into account equation (\ref{2.2}) which
relates the primary fields $\psi $, $\varphi $ and $p$. Introducing complex
Lagrange multiplier fields and requiring the action to be real, one arrives
at the following constraint term 
\begin{equation}
S_{c}=\int \left[ d^{2}z\right] \left[ \lambda \left( \partial \psi
-p\varphi \right) +\sigma \left( \overline{\partial }\varphi +p\psi \right)
+c.c.\right] \qquad  \label{3.2}
\end{equation}
which has to be added to the action (\ref{3.1}) 
(here and below $c.c.$ means complex conjugation). 
It is well known that, once
constraints are introduced into the generating functional of Green functions 
$Z=\int \left[ D\Pi \right] \exp (-S)$ , then the correct definition of
measure $\left[ D\Pi \right] $ requires the evaluation of the Faddeev-Popov
determinant. In our case fields are constrained by the Dirac equation (\ref
{2.2}), \textit{i.e.} 
\begin{equation}
L\left( 
\begin{array}{l}
\psi  \\ 
\varphi 
\end{array}
\right) =\left( 
\begin{array}{cc}
\partial & -p \\ 
p &\overline{\partial}
\end{array}
\right) \left( 
\begin{array}{l}
\psi  \\ 
\varphi 
\end{array}
\right) =0\qquad .  \label{3.3}
\end{equation}
To evaluate $\det (L)$ we will follow the heat kernel procedure (see [27],
[16]). The Faddeev-Popov action term is defined via 
\begin{equation}
S_{FP\text{ }}^{\prime }=-\log \left[ \det \,^{\prime }\left( L\right)
\right] =\left[ \frac{d}{ds}\zeta ^{\prime }\left( s|L\right) \right]
_{s=0}=\left[ \frac{1}{2}\frac{d}{ds}\zeta ^{\prime }\left( s|A\right)
\right] _{s=0}  \label{3.4}
\end{equation}
where $A=L^{\dagger}L=L L^{\dagger}$ and
\[
\zeta ^{\prime }\left( s|A\right) =\text{tr}^{\prime }\left[ \left(
A\right) ^{-s}\right] =\frac{1}{\Gamma \left( s\right) }\int_{0}^{\infty
}t^{s-1}\text{tr}^{\prime }\left[ \exp \left( -tA\right) \right] \qquad .
\]
Here the Riemann function $\xi $ is constructed from eigenvalues of the
operator $A$ and the prime means that the contribution of zero modes is
omitted. The heat kernel of $A$ is defined as [27]: 
\begin{equation}
\begin{array}{l}
K_{t}\left( z,z^{\prime }|A\right) =\left[ \exp \left( -tA\right)
\right] \left( z,z^{\prime }\right) \quad , \\ 
\partial _{t}K_{t}\left( z,z^{\prime }|A\right) +AK_{t}\left(
z,z^{\prime }|A\right) =0\quad , \\ 
\lim_{t\rightarrow 0^{+}}K_{t}\left( z,z^{\prime }|A\right) =\delta
\left( z,z^{\prime }\right) \qquad .
\end{array}
\label{3.5}
\end{equation}
The small $t$ expansion of $A$ looks like 
\begin{equation}
K_{t}\left( z,z^{\prime }|A\right) |_{z=z^{\prime }}\cong \frac{1}{8\pi t%
}\sum k_{n}\left( z\right) t^{n}  \label{3.6}
\end{equation}
where the $k_{n}\left( z\right) $ are matrix-valued functions. As a result,
one gets 
\begin{equation}
\zeta ^{\prime }\left( 0|A\right) =\frac{1}{4\pi }\int \left[
d^{2}z\right] \,
tr'[k_{1}\left( z\right)] \qquad ,\qquad \left[ \frac{d}{ds}\zeta
^{\prime }\left( s|A\right) \right] _{s=0}=\gamma \zeta ^{\prime }\left(
0|A\right)   \label{3.7}
\end{equation}
where $\gamma $ is the Euler-Mascheroni constant. Since for operator $A$
one has $tr'\left[ k_{1} \right]
=-2p^{2} $ [27], we finally get 
\begin{equation}
S_{FP}=-\frac{\gamma }{4\pi } \int \left[d^2z \right] p^2  
\label{sfp}
\end{equation}
Thus the effect of 
Faddeev-Popov determinant is reduced into a redefinition of the
 extrinsic coupling constant $\beta _{0}^{\prime }$. Such a new coupling
constant will be denoted as $\beta $ and we will assume 
$\beta > 0$. Therefore, the total action turns out
to be 
\begin{equation}
S_{T}=S_{NG}+S_{c}+S_{ext}  \label{act}
\end{equation}
where are $S_{NG}$, and $S_{c}$ are defined by (\ref{2.7}), (\ref{3.2}) and $%
S_{ext}=\beta \int \left[ d^{2}z\right] \,p^{2}$.

First, let us concentrate on the pure extrinsic action: $\alpha _{0}=0$. 
\textit{i.e.} 
\begin{equation}
S_{T}=S_{c}+S_{ext}\qquad .  \label{3.10}
\end{equation}
The corresponding action in one-loop approximation can be derived by
employing the background field method. So splitting all fields into slow and
fast components ($p=p_{0}+p_{1}$ etc.) and keeping only terms quadratic in
fast variables, we get the following one-loop action 
\begin{eqnarray}
\widetilde{S}^{(2)} &=&\int \left[ d^{2}z\right] \,[\lambda _{1}\left(
\partial \psi _{1}-p_{0}\varphi _{1}-p_{1}\varphi _{0}\right) +\sigma
_{1}\left( \overline{\partial }\varphi _{1}+p_{0}\psi _{1}+p_{1}\psi
_{0}\right)   \nonumber  \\
&&+p_{1}\left( \psi _{1}\sigma _{0}-\varphi _{1}\lambda _{0}\right)
+c.c.]+\beta \int \left[ d^{2}z\right] \,p_{1}^{2}\qquad .  \label{3.11}
\end{eqnarray}
In Fourier space we have $\widetilde{S}%
^{(2)}=\int \left[ d^{2}k\right] \,\widetilde{S}^{(2)}\left( k\right) $
where 
\begin{equation}
\begin{array}{lll}
\widetilde{S}^{(2)}\left( k\right) =& \beta \,\overline{p}\left( k\right)
\,p\left( k\right) +  \\
&+\left\{ \overline{\lambda }_{R}\left( k\right) \,
\left[ \frac{ \left( k+%
\overline{k}\right)}{4i} \psi _{R}\left( k\right) -\frac{\left( k-\overline{k%
}\right)}{4} \psi _{I}\left( k\right) -
p_{0}\varphi _{R}\left( k\right) -p\left(
k\right) \varphi _{0,R}\right] + \right. \\
&-\overline{\lambda }_{I}\left( k\right) \,\left[ \frac{\left( 
k- \overline{k}\right)}{4} \psi _{R}\left( k\right) +\frac{\left( k+%
\overline{k}\right)}{4i} \psi _{I}\left( k\right) -p_{0}\varphi _{I}\left(
k\right) -p\left( k\right) \varphi _{0,I}\right] +  \\
&+\overline{\sigma }_{R}\left( k\right) \,\left[ \frac{\left( k+%
\overline{k}\right)}{4i} \varphi _{R}\left( k\right) 
+\frac{\left( k- \overline{k}\right)}{4} 
\varphi _{I}\left( k\right) +p_{0}\psi _{R}\left( k\right)
+p\left( k\right) \psi _{0,R}\right] +  \\
&+\overline{\sigma }_{I}\left( k\right) \,\left[ \frac{\left( k-%
\overline{k}\right)}{4} \varphi _{R}\left( k\right) -\frac{\left( k+%
\overline{k}\right)}{4i} \varphi _{I}\left( k\right) -p_{0}\psi _{I}\left(
k\right) -p\left( k\right) \psi _{0,I}\right] +  \\
&
\left.
+\overline{p} \left( k \right) \left[ \sigma _{0,R}\psi _{R}\left( k\right)
-\sigma _{0,I}\psi _{I}\left( k\right) -\lambda _{0,R}\varphi _{R}\left(
k\right) +\lambda _{0,I}\varphi _{I}\left( k\right) \right] +c.c. 
\right\}  
\end{array}
\label{3.12}
\end{equation}
and $\left[ d^{2}k\right] =\left( i/2\right) \,dk\,d\overline{k}$. 
Here the subscripts $R$ and $I$ mean the real and imaginary
parts of complex fields involved in (\ref{3.11}) respectively 
and the fast field index $1$ is suppressed for simplicity.
In deriving (%
\ref{3.12}) we used the symmetrization procedure so that $\overline{S}%
^{\left( 2\right) }\left( k\right) =S^{\left( 2\right) }\left( -k\right) $ .
As usual, Fourier components $\,f\left( k\right) =\,f\left( k,\overline{k}%
\right) $ of fast fields are defined via 
\[
f\left( z,\overline{z} \right) =\frac{1}{2\pi}
\int \left[ d^{2}k \right] \,f\left( k\right)
\exp \left[ - \frac{i}{2}\left( k\overline{z}+\overline{k}z\right) \right] 
\]
where the integration is performed over the domain $\widetilde{\Lambda }%
<\left| k\right| <\Lambda $.

Exact propagators can be calculated using standard methods. 
They are presented in the Appendix. In particular, 
in the ultraviolet regime the two-point functions behave like
\[
\left\langle \overline{p}\left( k\right) \,p\left( k\right) \right\rangle
\simeq 2\beta ^{-1} \quad,
\]
\[
\left\langle \overline{\psi }_{R}\left( k\right) ,\lambda _{R}\left(
k\right) \right\rangle \,\,\,,\,\,\left\langle \overline{\psi }_{I}\left(
k\right) ,\lambda _{I}\left( k\right) \right\rangle \,\,\,,\,\,\left\langle 
\overline{\varphi }_{R}\left( k\right) ,\sigma _{R}\left( k\right)
\right\rangle \,\,,\,\,\left\langle \overline{\varphi }_{I}\left( k\right)
,\sigma _{R}\left( k\right) \right\rangle \propto \left| k\right| ^{-1}
\]
\[
\left\langle \overline{\psi }_{R}\left( k\right) ,\lambda _{R}\left(
k\right) \right\rangle \,\,\,,\,\,\left\langle \overline{\psi }_{I}\left(
k\right) ,\lambda _{R}\left( k\right) \right\rangle \,\,\,,\,\,\left\langle 
\overline{\varphi }_{R}\left( k\right) ,\sigma _{I}\left( k\right)
\right\rangle \,\,\,,\,\,\left\langle \overline{\varphi }_{I}\left( k\right)
,\sigma _{R}\left( k\right) \right\rangle \propto \left| k\right| ^{-1}\,\,
\]
and like $\left| k\right| ^{-2}$ in all the other cases.

\section{Perturbative evaluation of Nambu-Goto and spontaneous curvature
action}

Results obtained in previous section can be used for a perturbative analysis
of the intrinsic geometry term in the action (\ref{3.1}). At the one-loop
level the Nambu-Goto action reads 
\begin{eqnarray*}
S_{NG}^{(2)} =\int [d^2z] {\pounds}_{NG}^{(2)} &=&
2\alpha _{0}\int \left[ d^{2}z\right] \left[ \left( \left|
\psi _{0}\right| ^{2}+\left| \varphi _{0}\right| ^{2}\right) 
\left( 
\psi_R ^{2}+\psi_I^2 +\varphi_R^{2} +\varphi_I^2 \right) +\right. \\
&&\left.+2\left( \psi _{0,R}\psi _{R}+\psi _{0,I}\psi _{I}+\varphi
_{0,R}\varphi _{R}+\varphi _{0,R}\varphi _{R}\right) ^{2}\right]
\quad .
\end{eqnarray*}

The contribution to the classical Lagrangian from the 
intrinsic geometry term treated perturbatively is 
\[
\Delta S_{NG}^{(2)} = \int \left[ d^2z \right] 
<{\pounds}_{NG}^{(2)}> \qquad .  
\]
Using the propagators following from (\ref{3.12}), we get
\[
<{\pounds}_{NG}^{(2)}>
=\frac{\alpha _{0}}{ \pi^2}\left(
\left| \psi _{0}\right| ^{2}+\left| \varphi _{0}\right| ^{2}\right) ^{2}%
\int \left[ d^{2}k\right]
\left( \left| k\right| ^{2}+12p_{0}^{2}\right) D_P^{-1}
\qquad .  \label{4.1}
\]
Therefore, 
the counterterm to the classical Nambu-Goto action turns out to be 
\[
\Delta S_{NG}^{(2)}=
\frac{\alpha _{0}}{ \beta \left( \xi-1\right) }
I\left( \Lambda ,\widetilde{\Lambda },
p_{0}^{2},\xi,3 \right) 
\int \left[ d^2z \right]
\left( \left| \psi _{0}\right| ^{2}+\left|
\varphi _{0}\right| ^{2}\right) ^{2}
\]
where 
\[
I\left( \Lambda ,\widetilde{\Lambda },p_{0}^{2},\xi,a\right) =
\log \left[ 
\frac{\left( \left| k\right| ^{2}-4\xi p_{0}^{2}\right)^{\xi
+a}}{\left( \left| k\right| ^{2}-4p_{0}^{2}\right) ^{1+a}}\right]
_{\left| k\right| =\widetilde{\Lambda }}^{\left| k\right| =\Lambda }\qquad .
\]
and $\xi=\left( 1+2\beta _{0}^{\prime }/\beta \right)$. 
This counterterm gives rise to the renormalized Nambu-Goto action 
\[
\widetilde{S}_{NG}=
\widetilde{\alpha }\int \left[ d^{2}z\right] \left( \left| \psi
_{0}\right| ^{2}+\left| \varphi _{0}\right| ^{2}\right) ^{2}
\]
where $\widetilde{\alpha }=\alpha _{0}\left[ 1+\left( \pi \beta 
\right)^{-1}
\left( \xi -1\right) ^{-1} I\left( \Lambda ,\widetilde{\Lambda 
},p_{0}^{2},\xi,3 \right) \right] $ is the renormalized string tension.

Since the generalized Weierstrass representation allows to express
other action terms in simples forms, 
above perturbative approach can be successfully pursuit
in more general cases. In particular, we focus on a
term which is peculiar of surfaces in $\Bbb{R}^3$. It is the 
spontaneous curvature action defined as (see F. David in [3]): 
\[
S_{H}=\frac{\eta _{0}}{2}\int \left[ dS\right] \,H\qquad .
\]
In terms of the generalized Weierstrass representation, it results
(see eq. (\ref{2.4}))  
\begin{equation}
S_{H}=\eta _{0}\int \left[ d^{2}z\right] \,pu\qquad .  \label{sh}
\end{equation}
The one-loop spontaneous curvature action following from 
(\ref{sh}) reads 
\begin{equation}
\begin{array}{ll}
S_{H}^{(2)}=\int [d^2z] {\pounds}_H^{(2)} 
= & \eta _{0}\int \left[ d^{2}z\right] \,\left[ p_{0}\left( \left|
\psi \right| ^{2}+\left| \varphi \right| ^{2}\right) + 
\right. \nonumber \\
& \left. + 2p\left( \psi
_{R,0}\psi _{R}+\psi _{I,0}\psi _{I}+\varphi _{R,0}\varphi _{R}+\varphi
_{I,0}\varphi _{I}\right) \right] \quad .
\end{array}
\end{equation}
The counterterm to the classical spontaneous curvature action
looks like
\[
\Delta S_{H}^{(2)} =  \int [d^2z] <{\pounds}_H^{(2)}>
\]
where
\[
<{\pounds}_H^{(2)}>
=\frac{\eta _{0}}{ 2 \pi^2} p_{0}\left( \left|
\psi _{0}\right| ^{2}+\left| \varphi _{0}\right| ^{2}\right) 
\int \left[ d^{2}k\right] \left( 3\left|
k\right| ^{2}-4p_{0}^{2}\right) D_{P}^{-1}
\]
Therefore, the renormalized spontaneous curvature action reads 
\[
\widetilde{S}_{H}
=\widetilde{\eta }\int  \left[ d^{2}z\right] \,p_{0}\left( \left|
\psi _{0}\right| ^{2}+\left| \varphi _{0}\right| ^{2}\right) 
\]
where $
\widetilde{\eta }=\eta _{0}\left[ 1+ 3 \left( 2 \pi 
\beta \right)^{-1} \left( \xi -1\right)^{-1} 
I\left( \Lambda ,\widetilde{\Lambda },p_{0}^{2},\xi,
-\frac{1}{3}\right) \right] $ .

\section{Discussion}

In this paper we have calculated one-loop effects in the theory of a string
world-sheet conformally immersed in $\Bbb{R}^{3}$. We have considered the
complete Nambu-Goto-Polyakov action. Propagators for primary fields have
been calculated in the case of a pure extrinsic action (extrinsic Polyakov
action). In general, all of them have the following structure 
(see the Appendix) 
\begin{equation}
\frac{N_{P}}{D_{P}}  \label{n/p}
\end{equation}
where $N_{P}$'s are functions of slow fields and momentum and 
\[
D_{P}=\beta \left( \left| k\right| ^{2}-4p_{0}^{2}\right) \left( \left|
k\right| ^{2}-4\xi p_{0}^{2}\right) 
\]
where $\xi =\left( 1+2\beta _{0}^{\prime }/\beta \right) $. The
Faddeev-Popov determinant contributes to (\ref{n/p}) by mean of the
$\beta$. In our approach, 
we have a simple geometrical interpretation of the results since, 
in virtue of (\ref{2.4}), 
the $p_{0}$ field acts as a link between extrinsic and intrinsic geometry of
strings world-sheets.

The appearance of singularities which
depend on background geometry, as in (\ref{n/p}), seems to be completely new.
This is due to the fact that we have treated the full extrinsic
Polyakov action while
the other authors dealt principally with the kinetic part of one-loop action
[12] or by fixing minimal surface as background [13]. As a
result, these singularities did not show up in their formulae.

The singular behavior of propagators is naturally removed once an
infrared cut-off $\widetilde{\Lambda }$ is introduced. An infrared
cut-off constrained by the background geometry is commonly used in theory
of random surfaces.
Nevertheless, in our formalism the role of intrinsic geometry
is more transparent even in the pure extrinsic action case. 
The key point relies on the one-loop
approximation: quantum fluctuations of fields must have large momenta with
respect to the classical background ones. In our language this means that
infrared cut-off of the model $\widetilde{\Lambda }$ must be sufficiently
large with respect to $2p_{0}$, as it can be easily understood by rewriting
equation (\ref{2.2}) in the Fourier space. Therefore, inspite of the fact 
that the intrinsic
geometry does not play a role explicitly, it enters into the constraint for 
$\widetilde{\Lambda }$ by mean of (\ref{2.4}). In view of this, it
would be interesting to study how the Nambu-Goto and spontaneous curvature
actions contribute to propagators. This will be done in a separate paper.

Coming back to (\ref{n/p}) 
we note that, when $\xi $ is of the order of unity, the
points $\left| k_{1}\right| =2p_{0}$ and $\left| k_{2}\right| =\sqrt{4\xi
p_{0}^{2}}$ do not belong to the allowed spectrum of momenta and hence 
the two-points functions are well-defined. If $\xi >>1$, we can set 
$\widetilde{\Lambda }=\sqrt{4 \xi p_0^2} $. 

In the case of minimal surfaces $p_{0}\rightarrow 0$ and, consequently, the
formulae obtained for propagators are drastically simplified and hold
into a wide range of $\left| k\right| $. The infrared singularity is absent
provided that $\left| k\right| \geq \widetilde{\Lambda }>0$. The perturbative
analysis of the Nambu-Goto and the spontaneous curvature actions leads to a
logarithmic dependence on $\left| k\right| $ of the effective tension and
spontaneous curvature couplings according to

\begin{equation}
\alpha \left( \left| k\right| \right) \simeq \alpha _{0}\left[ 1+
\frac{2}{\beta \pi }\log 
\left( \frac{\Lambda }{\left| k\right| }\right) \right]
\qquad ,\qquad \eta \left( \left| k\right| \right) \simeq \eta _{0}
\left[ 1+
\frac{3}{ \pi \beta }\log \left( \frac{\Lambda }{\left| k\right| }\right)
\right] \qquad .  \label{runcou}
\end{equation}
The short wavelength regime takes place for such $k$ that $\left| k\right|
^{2}\gg 4\xi p_{0}^{2}$ . General formulae look similar to the minimal
surface case, the only difference is the value of the infrared
cut-off $\widetilde{\Lambda }$.

We would like also to mention that the propagators between the
Lagrange multiplier fields have the right sign, in contrast to the results
obtained in [12].

\hfill \hfill

\begin{center}
\textbf{APPENDIX}
\end{center}

\[
\left\langle \overline{p}\left( k\right) \,p\left( k\right) \right\rangle
=\left( \left| k\right| ^{2}-4p_{0}^{2}\right) \widetilde{D}_{P}^{-1}
\]
\[
\left\langle \overline{\psi }_{R}\left( k\right) \,\psi _{R}\left( k\right)
\right\rangle =2\left| 2p_{0}\func{Re}(\psi _{0})+i\func{Re}(k\varphi
_{0})\right| ^{2}D_{P}^{-1}
\]
\hfill 
\[
\left\langle \overline{\psi }_{I}\left( k\right) \,\psi _{I}\left( k\right)
\right\rangle =2\left| 2p_{0}\func{Im}(\psi _{0})+i\func{Im}(k\varphi
_{0})\right| ^{2}D_{P}^{-1}
\]
\hfill \hfill 
\[
\left\langle \overline{\varphi }_{R}\left( k\right) \,\varphi _{R}\left(
k\right) \right\rangle =2\left| 2p_{0}\func{Re}(\varphi _{0})+i\func{Re}(k%
\overline{\psi }_{0})\right| ^{2}D_{P}^{-1}
\]
\hfill \hfill 
\[
\left\langle \overline{\varphi }_{I}\left( k\right) \,\varphi _{I}\left(
k\right) \right\rangle =2\left| 2p_{0}\func{Im}(\varphi _{0})+i\func{Im}(k%
\overline{\psi }_{0})\right| ^{2}D_{P}^{-1}
\]
\hfill 
\[
\left\langle \overline{\lambda }_{R}\left( k\right) \,\lambda _{R}\left(
k\right) \right\rangle =2\left| 2p_{0}\func{Re}(\lambda _{0})+i\func{Re}%
(k\sigma _{0})\right| ^{2}D_{P}^{-1}
\]
\hfill 
\[
\left\langle \overline{\lambda }_{I}\left( k\right) \,\lambda _{I}\left(
k\right) \right\rangle =2\left| 2p_{0}\func{Im}(\lambda _{0})+i\func{Im}%
(k\sigma _{0})\right| ^{2}D_{P}^{-1}
\]
\hfill \hfill 
\[
\left\langle \overline{\sigma }_{R}\left( k\right) \,\sigma _{R}\left(
k\right) \right\rangle =2\left| 2p_{0}\func{Re}(\sigma _{0})+i\func{Re}(k%
\overline{\lambda }_{0})\right| ^{2}D_{P}^{-1}
\]
\hfill 
\[
\left\langle \overline{\sigma }_{I}\left( k\right) \,\sigma _{I}\left(
k\right) \right\rangle =2\left| 2p_{0}\func{Im}(\sigma _{0})+i\func{Im}(k%
\overline{\lambda }_{0})\right| ^{2}D_{P}^{-1}
\]
\hfill \hfill 
\[
\left\langle \overline{p}\left( k\right) ,\psi _{R}\left( k\right)
\right\rangle =2 \left[ 2p_{0}\func{Re}(\psi _{0})+i\func{Re}(k\varphi
_{0})\right] \widetilde{D}_{P}^{-1}
\]
\hfill 
\[
\left\langle \overline{p}\left( k\right) ,\psi _{I}\left( k\right)
\right\rangle =2 \left[ 2p_{0}\func{Im}(\psi _{0})+i\func{Im}(k\varphi
_{0})\right] \widetilde{D}_{P}^{-1}
\]
\hfill \hfill 
\[
\left\langle \overline{p}\left( k\right) ,\varphi _{R}\left( k\right)
\right\rangle =2
\left[ 2p_{0}\func{Re}(\varphi _{0})-i\func{Re}(k\overline{%
\psi }_{0})\right] \widetilde{D}_{P}^{-1}
\]
\hfill 
\[
\left\langle \overline{p}\left( k\right) ,\varphi _{I}\left( k\right)
\right\rangle =2
\left[ 2p_{0}\func{Im}(\varphi _{0})+i\func{Im}(k\overline{%
\psi }_{0})\right] \widetilde{D}_{P}^{-1}
\]
\hfill \hfill 
\[
\left\langle \overline{p}\left( k\right) ,\lambda _{R}\left( k\right)
\right\rangle =2
\left[ 2p_{0}\func{Re}(\lambda _{0})+i\func{Re}(k\sigma
_{0})\right] \widetilde{D}_{P}^{-1}
\]
\hfill 
\[
\left\langle \overline{p}\left( k\right) ,\lambda _{I}\left( k\right)
\right\rangle =2
\left[ 2p_{0}\func{Im}(\lambda _{0})+i\func{Im}(k\sigma
_{0})\right] \widetilde{D}_{P}^{-1}
\]
\hfill \hfill 
\[
\left\langle \overline{p}\left( k\right) ,\sigma _{R}\left( k\right)
\right\rangle =2
\left[ 2p_{0}\func{Re}(\sigma _{0})-i\func{Re}(k\overline{%
\lambda }_{0})\right] \widetilde{D}_{P}^{-1}
\]
\hfill 
\[
\left\langle \overline{p}\left( k\right) ,\sigma _{I}\left( k\right)
\right\rangle =2
\left[ 2p_{0}\func{Im}(\sigma _{0})+i\func{Im}(k\overline{%
\lambda }_{0})\right] \widetilde{D}_{P}^{-1}
\]
\hfill \hfill 
\[
\left\langle \overline{\psi }_{R}\left( k\right) ,\psi _{I}\left( k\right)
\right\rangle =2\left( 2p_{0}\psi _{0R}-i\func{Re}\left[ k\varphi
_{0}\right] \right) \left( 2p_{0}\psi _{0I}+i\func{Im}\left[ k\varphi
_{0}\right] \right) D_{P}^{-1}
\]
\hfill 
\[
\left\langle \overline{\psi }_{R}\left( k\right) ,\varphi _{R}\left(
k\right) \right\rangle =2\left( 2p_{0}\psi _{0R}-i\func{Re}\left[ k\varphi
_{0}\right] \right) \left( 2p_{0}\varphi _{0R}-i\func{Re}\left[ k\overline{%
\psi }_{0}\right] \right) D_{P}^{-1}
\]
\hfill \hfill 
\[
\left\langle \overline{\psi }_{R}\left( k\right) ,\varphi _{I}\left(
k\right) \right\rangle =2\left( 2p_{0}\psi _{0R}-i\func{Re}\left[ k\varphi
_{0}\right] \right) \left( 2p_{0}\varphi _{0I}+i\func{Im}\left[ k\overline{%
\psi }_{0}\right] \right) D_{P}^{-1}
\]
\hfill \hfill 
\[
\left\langle \overline{\psi }_{R}\left( k\right) ,\lambda _{R}\left(
k\right) \right\rangle =\left[ \frac{(k+\overline{k})}{4i}\widetilde{D}%
_{P}+2\left( 2p_{0}\psi _{0R}-i\func{Re}\left[ k\varphi _{0}\right] \right)
\left( 2p_{0}\lambda _{0R}+i\func{Re}\left[ k\sigma _{0}\right] \right)
\right] D_{P}^{-1}
\]
\hfill 
\[
\left\langle \overline{\psi }_{R}\left( k\right) ,\lambda _{I}\left(
k\right) \right\rangle =\left[ -\frac{(k-\overline{k})}{4}\widetilde{D}%
_{P}+2\left( 2p_{0}\psi _{0R}-i\func{Re}\left[ k\varphi _{0}\right] \right)
\left( 2p_{0}\lambda _{0I}+i\func{Im}\left[ k\sigma _{0}\right] \right)
\right] D_{P}^{-1}
\]
\hfill \hfill 
\[
\left\langle \overline{\psi }_{R}\left( k\right) ,\sigma _{R}\left( k\right)
\right\rangle =\left[ -p_{0}\widetilde{D}_{P}+4\left( 2p_{0}\psi _{0R}-i%
\func{Re}\left[ k\varphi _{0}\right] \right) \left( 2p_{0}\sigma _{0R}-i%
\func{Re}\left[ k\overline{\lambda }_{0}\right] \right) \right] D_{P}^{-1}
\]
\hfill \hfill 
\[
\left\langle \overline{\psi }_{R}\left( k\right) ,\sigma _{I}\left( k\right)
\right\rangle =2\left( 2p_{0}\psi _{0R}-i\func{Re}\left[ k\varphi
_{0}\right] \right) \left( 2p_{0}\sigma _{0I}+i\func{Re}\left[ k\overline{%
\lambda }_{0}\right] \right) D_{P}^{-1}
\]
\hfill \hfill 
\[
\left\langle \overline{\psi }_{I}\left( k\right) ,\varphi _{R}\left(
k\right) \right\rangle =2\left( 2p_{0}\psi _{0R}-i\func{Im}\left[ k\varphi
_{0}\right] \right) \left( 2p_{0}\varphi _{0R}-i\func{Re}\left[ k\overline{%
\psi }_{0}\right] \right) D_{P}^{-1}
\]
\hfill \hfill 
\[
\left\langle \overline{\psi }_{I}\left( k\right) ,\varphi _{I}\left(
k\right) \right\rangle =2\left( 2p_{0}\psi _{0R}-i\func{Im}\left[ k\varphi
_{0}\right] \right) \left( 2p_{0}\varphi _{0I}+i\func{Im}\left[ k\overline{%
\psi }_{0}\right] \right) D_{P}^{-1}
\]
\hfill \hfill 
\[
\left\langle \overline{\psi }_{I}\left( k\right) ,\lambda _{R}\left(
k\right) \right\rangle =\left[ -\frac{(k-\overline{k})}{4}\widetilde{D}%
_{P}+4\left( 2p_{0}\psi _{0I}-i\func{Im}\left[ k\varphi _{0}\right] \right)
\left( 2p_{0}\lambda _{0R}+i\func{Re}\left[ k\sigma _{0}\right] \right)
\right] D_{P}^{-1}
\]
\hfill \hfill 
\[
\left\langle \overline{\psi }_{I}\left( k\right) ,\lambda _{I}\left(
k\right) \right\rangle =\left[ i\frac{(k+\overline{k})}{4}\widetilde{D}%
_{P}+4\left( 2p_{0}\psi _{0I}-i\func{Im}\left[ k\varphi _{0}\right] \right)
\left( 2p_{0}\lambda _{0I}+i\func{Im}\left[ k\sigma _{0}\right] \right)
\right] D_{P}^{-1}
\]
\hfill \hfill 
\[
\left\langle \overline{\psi }_{I}\left( k\right) ,\sigma _{R}\left( k\right)
\right\rangle =2\left( 2p_{0}\psi _{0I}-i\func{Im}\left[ k\varphi
_{0}\right] \right) \left( 2p_{0}\sigma _{0R}-i\func{Re}\left[ k\overline{%
\lambda }_{0}\right] \right) D_{P}^{-1}
\]
\hfill \hfill 
\[
\left\langle \overline{\psi }_{I}\left( k\right) ,\sigma _{I}\left( k\right)
\right\rangle =\left[ p_{0}\widetilde{D}_{P}+4\left( 2p_{0}\psi _{0I}-i%
\func{Im}\left[ k\varphi _{0}\right] \right) \left( 2p_{0}\sigma _{0I}+i%
\func{Im}\left[ k\overline{\lambda }_{0}\right] \right) \right] D_{P}^{-1}
\]
\hfill \hfill 
\[
\left\langle \overline{\varphi }_{R}\left( k\right) ,\varphi _{I}\left(
k\right) \right\rangle =2\left( 2p_{0}\varphi _{0R}+i\func{Re}\left[ k%
\overline{\psi }_{0}\right] \right) \left( 2p_{0}\varphi _{0I}+i\func{Im}%
\left[ k\overline{\psi }_{0}\right] \right) D_{P}^{-1}
\]
\hfill \hfill 
\[
\left\langle \overline{\varphi }_{R}\left( k\right) ,\lambda _{R}\left(
k\right) \right\rangle =\left[ p_{0}\widetilde{D}_{P}+4\left( 2p_{0}\varphi
_{0R}+i\func{Re}\left[ k\overline{\psi }_{0}\right] \right) \left(
2p_{0}\lambda _{0R}+i\func{Re}\left[ k\sigma _{0}\right] \right) \right]
D_{P}^{-1}
\]
\hfill \hfill 
\[
\left\langle \overline{\varphi }_{R}\left( k\right) ,\lambda _{I}\left(
k\right) \right\rangle =2\left( 2p_{0}\varphi _{0R}+i\func{Re}\left[ k%
\overline{\psi }_{0}\right] \right) \left( 2p_{0}\lambda _{0I}+i\func{Im}%
\left[ k\sigma _{0}\right] \right) D_{P}^{-1}
\]
\hfill \hfill 
\[
\left\langle \overline{\varphi }_{R}\left( k\right) ,\sigma _{R}\left(
k\right) \right\rangle =\left[ \frac{(k+\overline{k})}{4i}\widetilde{D}%
_{P}+4\left( 2p_{0}\varphi _{0R}+i\func{Re}\left[ k\overline{\psi }%
_{0}\right] \right) \left( 2p_{0}\sigma _{0R}-i\func{Re}\left[ k\overline{%
\lambda }_{0}\right] \right) \right] D_{P}^{-1}
\]
\hfill \hfill 
\[
\left\langle \overline{\varphi }_{R}\left( k\right) ,\sigma _{I}\left(
k\right) \right\rangle =\left[ \frac{(k-\overline{k})}{4}\widetilde{D}%
_{P}+4\left( 2p_{0}\varphi _{0R}+i\func{Re}\left[ k\overline{\psi }%
_{0}\right] \right) \left( 2p_{0}\sigma _{0I}+i\func{Im}\left[ k\overline{%
\lambda }_{0}\right] \right) \right] D_{P}^{-1}
\]
\hfill \hfill 
\[
\left\langle \overline{\varphi }_{I}\left( k\right) ,\lambda _{R}\left(
k\right) \right\rangle =2\left( 2p_{0}\varphi _{0I}-i\func{Im}\left[ k%
\overline{\psi }_{0}\right] \right) \left( 2p_{0}\lambda _{0R}+i\func{Re}%
\left[ k\sigma _{0}\right] \right) D_{P}^{-1}
\]
\hfill \hfill 
\[
\left\langle \overline{\varphi }_{I}\left( k\right) ,\lambda _{I}\left(
k\right) \right\rangle =\left[ -p_{0}\widetilde{D}_{P}+4\left(
2p_{0}\varphi _{0I}-i\func{Im}\left[ k\overline{\psi }_{0}\right] \right)
\left( 2p_{0}\lambda _{0I}+i\func{Im}\left[ k\sigma _{0}\right] \right)
\right] D_{P}^{-1}
\]
\hfill \hfill 
\[
\left\langle \overline{\varphi }_{I}\left( k\right) ,\sigma _{R}\left(
k\right) \right\rangle =\left[ \frac{k-\overline{k}}{4}\widetilde{D}%
_{P}+4\left( 2p_{0}\varphi _{0I}-i\func{Im}\left[ k\overline{\psi }%
_{0}\right] \right) \left( 2p_{0}\sigma _{0R}-i\func{Re}\left[ k\overline{%
\lambda }_{0}\right] \right) \right] D_{P}^{-1}
\]
\hfill \hfill 
\[
\left\langle \overline{\varphi }_{I}\left( k\right) ,\sigma _{I}\left(
k\right) \right\rangle =\left[ i\frac{(k+\overline{k})}{4}\widetilde{D}%
_{P}+4\left( 2p_{0}\varphi _{0I}-i\func{Im}\left[ k\overline{\psi }%
_{0}\right] \right) \left( 2p_{0}\sigma _{0I}+i\func{Im}\left[ k\overline{%
\lambda }_{0}\right] \right) \right] D_{P}^{-1}
\]
\hfill \hfill 
\[
\left\langle \overline{\lambda }_{R}\left( k\right) ,\lambda _{I}\left(
k\right) \right\rangle =2\left( 2p_{0}\lambda _{0R}-i\func{Re}\left[ k\sigma
_{0}\right] \right) \left( 2p_{0}\lambda _{0I}+\func{Im}\left[ k\sigma
_{0}\right] \right) D_{P}^{-1}
\]
\hfill \hfill 
\[
\left\langle \overline{\lambda }_{R}\left( k\right) ,\sigma _{R}\left(
k\right) \right\rangle =2\left( 2p_{0}\lambda _{0R}-i\func{Re}\left[ k\sigma
_{0}\right] \right) \left( 2p_{0}\sigma _{0R}-i\func{Re}\left[ k\overline{%
\lambda }_{0}\right] \right) D_{P}^{-1}
\]
\hfill \hfill 
\[
\left\langle \overline{\lambda }_{R}\left( k\right) ,\sigma _{I}\left(
k\right) \right\rangle =2\left( 2p_{0}\lambda _{0R}-i\func{Re}\left[ k\sigma
_{0}\right] \right) \left( 2p_{0}\sigma _{0I}+i\func{Im}\left[ k\overline{%
\lambda }_{0}\right] \right) D_{P}^{-1}
\]
\hfill \hfill 
\[
\left\langle \overline{\lambda }_{I}\left( k\right) ,\sigma _{R}\left(
k\right) \right\rangle =2\left( 2p_{0}\lambda _{0I}-i\func{Im}\left[ k\sigma
_{0}\right] \right) \left( 2p_{0}\sigma _{0R}-i\func{Re}\left[ k\overline{%
\lambda }_{0}\right] \right) D_{P}^{-1}
\]
\hfill \hfill 
\[
\left\langle \overline{\lambda }_{I}\left( k\right) ,\sigma _{I}\left(
k\right) \right\rangle =2\left( 2p_{0}\lambda _{0I}-i\func{Im}\left[ k%
\overline{\lambda }_{0}\right] \right) \left( 2p_{0}\sigma _{0I}+i\func{Im}%
\left[ k\overline{\lambda }_{0}\right] \right) D_{P}^{-1}
\]
\hfill \hfill 
\[
\left\langle \overline{\sigma }_{R}\left( k\right) ,\sigma _{I}\left(
k\right) \right\rangle =2\left( 2p_{0}\sigma _{0R}+i\func{Re}\left[ k%
\overline{\lambda }_{0}\right] \right) \left( 2p_{0}\sigma _{0I}+i\func{Im}%
\left[ k\overline{\lambda }_{0}\right] \right) D_{P}^{-1}
\]
where 
\[
D_{P}=\left( \left| k\right| ^{2}-4p_{0}^{2}\right) \left[ \beta \left(
\left| k\right| ^{2}-4p_{0}^{2}\right) -8p_{0}\func{Re}(\lambda _{0}\varphi
_{0}-\sigma _{0}\psi _{0})\right] 
\]
\[
\widetilde{D}_{P}=2\left( \left| k\right| ^{2}-4p_{0}^{2}\right) ^{-1}D_{P}
\]
By using the classical equation of motion $2\beta _{0}^{\prime }p_{0}=2\func{%
Re}(\lambda _{0}\varphi _{0}-\sigma _{0}\psi _{0})$, $D_{P}$ and 
$\widetilde{%
D}_{P}$ can be expressed in terms of the only momentum and $p_{0}$: 
\[
D_{P}=\beta \left( \left| k\right| ^{2}-4p_{0}^{2}\right) \left( \left|
k\right| ^{2}-4\xi p_{0}^{2}\right) \quad ,\quad \widetilde{D}_{P}=2\beta
\left( \left| k\right| ^{2}-4\xi p_{0}^{2}\right) 
\]
where $\xi =(1+2\beta _{0}^{\prime }/\beta )$.

\hfill \hfill

\begin{center}
\textbf{REFERENCES}
\end{center}

\begin{enumerate}
\item  A. M. Polyakov, Nucl. Phys., \textbf{B268}, 406-412, (1986).

\item  A. M. Polyakov, \textit{Gauge fields and strings}, Harwood Academic
Publishers, London, 1987.

\item  D. Nelson, T. Piran and S. Weinberg Eds., \textit{Statistical
Mechanics of Membranes and surfaces}, World Scientific, Singapore, 1989.

\item  F. David, P. Ginsparg and Y. Zinn-Justin Eds., \textit{Fluctuating
Geometries in Statistical Machanics and Field Theory}, Elsevier, Amsterdam,
1996.

\item  A. A. Belavin and V. G. Knizhnik, Phys. Lett, \textbf{B168}, 201,
(1986).

\item  A. A. Belavin, V. G. Knizhnik, A. Morozov and A. Perelomov, Phys.
Lett, \textbf{B177}, 324, (1986).

\item  G. M. Sotkov, M. Stanishkov and C-J. Zhu, Nucl. Phys., \textbf{B356},
(1991), 245.

\item  T. L. Curtright, G. I. Chandour and C. K. Zachos, Phys. Rev. \textbf{%
D34}, 3811, (1986).

\item  E. Braaten and C. K. Zachos, Phys. Rev., \textbf{D35}, 1512 (1987).

\item  H. Kleinert, Phys. Lett., \textbf{B17}4, 335, (1986).

\item  F. David and E. Guitter, Nucl. Phys., \textbf{B295}, 332, (1988).

\item  K. S. Viswanathan, R. Parthasarathy and D. Kay, Ann. Phys. \textbf{206%
}, 237-254, (1991).

\item  K. S. Viswanathan and R. Parthasarathy, Phys. Rev., \textbf{D51},
5830-5838, (1991).

\item  J. Ambj\o rn, \textit{Quantum Geometry: a statistical field approach}%
, Cambridge Univ. Press, Cambridge, (1997).

\item  B. G. Konopelchenko, Stud. Appl. Math. \textbf{96}, 9, (1996).

\item  S. Matsutani, \textit{Immersion anomaly of Dirac operator on surface
in} $\Bbb{R}^{3}$, preprint, physics/9707003.

\item  S. Matsutani, J. Phys. A: Math. Gen., \textbf{31}, (1998), 3595.

\item  I. Taimanov, Trans. Amer. Math. Soc., Ser. 2, \textbf{179}, 133-159,
(1997).

\item  K. Kenmotsu, Math. Ann., \textbf{245}, (1979), 89.

\item  R. Carroll and B. G. Konopelchenko, Int. J. Mod. Phys. \textbf{A11},
1183, (1996).

\item  T. .J. Willmore, \textit{Total Curvature in Riemannian Geometry},
Ellis Horwood, New York, 1982.

\item  B. G. Konopelchenko and I. Taimanov, preprint N. 187, Univ. Bochum,
(1995).

\item  P.G. Grinevich and M.V. Schmidt, preprint SFB288 N 291, TU-Berlin,
1997; Journal of Geometry and Physics (to appear).

\item  I. Taimanov, Uspechi Mat. Nauk, \textbf{52}, N 6, (1997), 187-188.

\item  B. G. Konopelchenko, Phys. Lett. \textbf{B414}, (1997), 58.

\item  B. G. Konopelchenko and G. Landolfi, Mod. Phys. Lett. \textbf{A12},
3161-3168, (1997).

\item  E. Elizalde, S.D. Odintosov, A. Romeo, A.A. Bytsenko, S. Zerbini, 
\textit{Zeta regularization techniques with applications}, World
Scientific, Singapore, 1994.
\end{enumerate}

\end{document}